# Investigating Anomalous Thermal Expansion of Copper Halides by Inelastic Neutron Scattering and Ab-inito Phonon Calculations

Abhijith M. Gopakumar[a], M. K. Gupta[b], R. Mittal[b], S. Rols[c] and S. L. Chaplot[b]
[a]Department of Physics, Indian Institute of Technology, Guwahati 781039
[b]Solid State Physics Division, Bhabha Atomic Research Centre, Mumbai 400085
[c]Institut Laue-Langevin, 71 Avenue des Martyrs, CS 20156, 38042 Grenoble Cedex 9, France

We investigate detailed lattice dynamics of copper halides CuX (X=Cl, Br, I) using neutron inelastic scattering measurements and ab-initio calculations aimed at a comparative study of their thermal expansion behavior. We identify the low energy phonons which soften with pressure and are responsible for negative thermal expansion. The eigenvector analysis of these modes suggests that softening of the transverse-acoustic modes would lead to NTE in these compounds. The calculations are in very good agreement with our measurements of phonon spectra and thermal expansion behavior as reported in the literature. Our calculations at high pressure further reveal that large difference in negative thermal expansion behavior in these compounds is associated with the difference of the unit cell volume.

**PACS**: 78.70.Nx, 63.20.-e, 65.40.-b, 63.20.dk

**Keywords:** Inelastic neutron scattering, Phonons in crystal lattice, Thermal properties of crystalline solids, First-principles theory

## I. INTRODUCTION

The research on materials with anomalous thermal expansion has become a popular area of interest in material science during past three decades[1-12]. The main reason for this increase in popularity is the potential usage of these compounds in various industrial applications. Designing an opto-electronic device requires the knowledge on their expansion with temperature. The volume expansion plays a major role in performance and thus the expansion should be tuned as per specific requirements. As there is no single element with zero thermal expansion coefficient, one has to depend on composite materials. This paved the way for a new field of composites – the thermal expansion compensators[13-15]. The zero expansion compositions require a material with negative thermal Expansion (NTE) and another material with positive thermal expansion (PTE). Large numbers of studies are being carried out to understand the phenomenon of NTE, which were useful in understanding the microscopic origin of mechanism in many compounds[16-22] etc. However, mechanism of NTE in various compounds is still ambiguous.

Some of the newly found NTE compounds have very high value of linear NTE coefficient even at room temperature. For example, the compounds like $Cd(CN)_2$ ($-20.4 \times 10^{-6}$ $K^{-1}$), $Zn(CN)_2$ ($-16.9 \times 10^{-6}$ $K^{-1}$), $CaZrF_6$ ($-13 \times 10^{-6}$ $K^{-1}$) and $CaHfF_6$ ($-14 \times 10^{-6}$ $K^{-1}$) have large NTE coefficients. A few mechanisms based on rigid unit model (RUM)[23, 24] and tent model[25, 26] have been proposed for explaining the NTE in various compounds. The models were successful in explaining NTE in many framework compounds[27, 28], however they were not found to explain the thermal expansion behaviour in many other cases[23, 29].

The thermal expansion behaviour of many compounds exhibiting NTE characteristics[30-35] have been determined mainly using diffraction techniques. The experimental thermal expansion studies of isostructural cuprous halides, CuX (X=Cl, Br, I) have been reported in the literature[32-35]. However the mechanism of NTE in these compounds is not understood. In this work, we have studied the phonon spectra of cuprous halides to explain their anomalous thermal expansion behavior. The three compounds, CuCl, CuBr, CuI, fall near the boundary between ionic and covalent[36]. In cuprous halides, the filled and interacting d orbital's provide surplus number of electrons while bonding. Cuprous halides form zinc blend type fcc crystal lattice shown in Fig. 1 (space group: F-43m). The size of anion is around twice of that of cation in cuprous halides and the cation is positioned at the centre of tetrahedral voids formed by anions. The natures of atomic vibrations in all these three halides are strongly anharmonic and thus their properties differ much from other isostructural compounds. So considering the similarities among these three compounds, we aim to derive the factors that would control the thermal expansion.

## II. EXPERIMENTAL DETAILS

The polycrystalline samples of CuX were purchased from Sigma Aldrich. The inelastic neutron scattering experiments on CuX (X=Cl, Br and I) were carried out using the IN4C spectrometers at the Institut Laue Langevin (ILL), France. The spectrometer is based on the time-of-flight technique and is equipped with a large detector bank covering a wide range of about $10^o$ to $110^o$ of scattering angle. Since these samples are hygroscopic, they were heated to 450 K to remove water absorbed by the sample. The samples were then cooled and the inelastic neutron scattering measurements are performed at above room temperature at 373 K, 473 K and 573 K respectively. For these measurements we have used an incident neutron wavelength of 2.4 Å (14.2 meV) in neutron energy gain setup. In the incoherent one-phonon approximation, the measured scattering function $S(Q,E)$, as observed in the neutron experiments, is related [37] to the phonon density of states $g^{(n)}(E)$ as follows:

$$g^{(n)}(E) = A < \frac{e^{2W(Q)}}{Q^2} \frac{E}{n(E,T) + \frac{1}{2} \pm \frac{1}{2}} S(Q,E) > \qquad (1)$$

$$g^n(E) = B \sum_k \{\frac{4\pi b_k^2}{m_k}\} g_k(E) \qquad (2)$$

where the + or − signs correspond to energy loss or gain of the neutrons respectively and where $n(E,T) = \left[\exp(E/k_BT) - 1\right]^{-1}$. $A$ and $B$ are normalization constants and $b_k$, $m_k$, and $g_k(E)$ are, respectively, the neutron scattering length, mass, and partial density of states of the $k^{th}$ atom in the unit cell. The quantity between <> represents suitable average over all $Q$ values at a given energy. $2W(Q)$ is the Debye-Waller factor averaged over all the atoms. The weighting factors $\frac{4\pi b_k^2}{m_k}$ for various atoms in the units of barns/amu are: 0.1264, 0.4739, 0.0738 and 0.0300 for Cu, Cl, Br and I respectively. The values of neutron scattering lengths for various atoms can be found from Ref[38].

## III. COMPUTATIONAL DETAILS

The calculations were performed using ab-intio density functional theory (DFT)[39, 40] method implemented in Vienna ab-initio simulation package (VASP)[41-44]. Projected Augmented Wave (PAW) potentials, included in VASP package, with Generalised Gradient Approximation (GGA) [45-47] were used for calculations. PAW method combines both pseudopotential and all electron methods in an optimized manner. PAW creates pseudopotentials that adjust to the instantaneous electronic structure and charge transferability problems of pseudopotential method are under control [48, 49]. The GGA was formulated by the Perdew–Burke–Ernzerhof (PBE) density functional. The valence electronic configurations of Cu, Cl, Br and I, as used in calculations for pseudo-potential generation are $3d^{10} 4s^1$, $3s^2 3p^5$, $3d^{10} 4s^2 4p^5$ and $4d^{10} 5s^2 5p^5$ respectively.

A plane wave kinetic energy cut off of 480 eV was used for electronic structure calculations of CuX (X=Cl,Br, I) Brillouin zone integrations were sampled using a Monkhorst-pack [50] generated 8×8×8 k-point mesh. The convergence criteria for total energy and ionic forces were set to be $10^{-8}$ and $10^{-5}$ respectively. The phonon frequencies have been calculated using PHONON software package[51] with a 2×2×2 super cell. Hellman Feynman forces were calculated using finite displacement method (with a displacement of 0.03 Angstroms) for obtaining phonon spectra. The optimized unit cell parameters are in good agreement with the previous reported values.

It is well known that thermal expansion in insulators originates from anharmonicity of the lattice vibrations[52-55]. The calculation of thermal expansion involves the calculation of the Grüneisen parameters of phonons. The phonon bandstructure in the entire Brillouin zone and its volume dependence under the quasiharmonic approximation were obtained. In quasiharmonic approximation, volume thermal expansion coefficient is given by

$$\alpha_V(T) = \frac{1}{BV} \sum_i \Gamma_i C_{V_i}(T) \quad (1)$$

Where $\Gamma_i$ ( $=-\partial lnE_i/\partial lnV=-B(\partial lnE_i/\partial lnP)$ ) and $C_{vi}$ are the mode Grüneisen parameter and specific heat due to the i$^{th}$ phonon at temperature T respectively, $E_i$ is phonon energy, V is the volume of the unit cell and B is the bulk modulus. Using this relation, we have calculated the volume thermal expansion from the pressure dependence of phonon spectra. Since $C_{Vi}(T)$ is positive for all modes at all temperatures, the nature of thermal expansion coefficient being positive or negative is only governed by the Grüneisen parameters of phonon modes.

## IV. RESULTS AND DISCUSSION

### A. Temperature Dependence of Phonon Spectra

The measured inelastic neutron scattering spectra for CuX (X= Cl, Br and I) at 373 K, 473 K and 573 K is shown in Fig 2. The lowest energy peak in all the three halides is at about 4-5 meV. This peak may be due to transverse acoustic modes. The highest energy peak which corresponds to the optic modes of Cu-X, occur at different energies of at about 27 meV, 19 meV and 16 meV for CuCl, CuBr and CuI respectively. The peak around 4-5 meV shows significant temperature dependence. In particular, in case of CuCl, this peak reduces in intensity and broadens, which indicates significant anharmonic behaviour. The effect is less pronounced in case of CuI.

The neutron spectra of the CuX have been interpreted by calculating the phonon spectra as well as partial contribution of various atoms to the total phonon spectra. The partial contributions to the various atoms have been estimated by atomic projections of the one-phonon eigenvectors. The calculated partial density of states is shown in Fig 3. The contributions due to both the Cu and X atoms extend over whole spectral range. In case of CuCl, the Cl (35.45 amu) has smaller mass in comparison to the Cu atom (63.55 amu), while in CuBr, the mass of Br (79.90 amu) is comparable to that of Cu, and I (126.90 amu) atoms are relatively heavier in comparison to Cu in CuI. We observe (Fig. 3) that in all three compounds the lowest peak in the phonon spectra at about 4-5 meV has contribution from both the Cu and X (=Cl, Br and I) atoms. We expect that contribution from the X (=Cl, Br and I) atoms would follow the mass consideration. However we find that the lowest peak in the partial contribution from X atoms in all the three halides is at nearly same energy of about 4-5 meV. This indicates that there is significant difference in bonding which compensates the mass effect.

The calculated partial density of states shows that the contribution from the Cu atom is very much different in various halides. At low energy below 10 meV, the vibrational contribution of Cu atoms to the total density of states in CuCl is largest while it is least in CuI. However, contribution below 10 meV from the halide atoms (X=Cl in CuCl) in the spectra is least in CuCl and largest in CuI. The energy range of the Cu-X optic modes in these compounds simply follows the reduce mass and bond length considerations. The smallest Cu-X mass and the bond length in CuCl results in the spectral range of the optic modes to extend up to 30 meV while the range of optic modes in CuBr and CuI is up to 20 meV and 18 meV respectively.

The comparison between measured (373 K) and calculated phonon spectra is shown in Fig 4. The experimental phonon spectrum in CuCl shows peaks centered at 4, 8, 12 and 28 meV. However, the calculated one-phonon spectrum shows three peak structures. We note that the experimental spectrum includes contributions from multiphonon. The additional peak in the measured spectrum is at 8 meV, which may arise due to the multiphonon contribution. In order to compare the calculated and the measured spectra, the multiphonon contribution has been estimated following the Sjolander formalism[56]. The total spectrum, which contain contribution from both the one-phonon and multiphonon parts, are compared (Fig 4) with the experimental spectra in all three halides. The comparison of the one-phonon and the total spectrum including multiphonon clearly shows that the peak in the experimental spectra at about 8-10 meV arises due to the contribution from the multiphonon in all the three compounds. Further we have computed the specific heat using the calculated phonon spectra in all three compounds. The excellent agreement (Fig. 5) between the calculations and experimental data[57, 58] further validates our calculations.

**B. Phonon Dispersion Relation**

We have computed phonon dispersion relation of all these three cuprous halides. In addition to that, we have computed the Born effective charges and static dielectric constants using linear response methods implemented in VASP. The Born effective charges and dielectric constants are used to calculate the longitudinal optical and transverse optical (LO-TO) splitting of the optic modes in the phonon spectra. In Fig. 6 we have compared the calculated and experimental dispersion relation[57, 58] of all the three CuX compounds. It can be seen that at low energies up to 15 meV there is an excellent agreement between calculations and measurement. We find that in both the Cl and I compounds the calculated longitudinal optic branch near the zone-centre is found to be underestimated, while the agreement for the same branch is found to be good in CuBr.

Although all the three compounds crystallize in the same space group (F-43m), the unit cell parameters show large variation. This would result in large variation in the bond lengths The energy range of the acoustic modes is found to be nearly same (up to about 15 meV) in all the three compounds. However there exists significant difference in the magnitude of the energy range of the Cu-X optic modes in all these three compounds. The optic modes in the Cl compound show large dispersion while they are almost flat in the I compound. The LO-TO splitting of the modes is found to be largest in CuCl and least in CuI.

**C. Thermodynamic Behavior of CuX (X=Cl, Br, I)**

The calculated and experimental values of lattice parameters[59] bulk moduli (B) are given in Table II. Experimental values were taken from literature [60]. The calculated values using the GGA are overestimated as compared with the experimental ones. The specific heat ($C_p$) are calculated from the phonon density of states and compared with the experimental results. The calculated $C_p$ values (Fig. 6) of all the three compounds agree well with the experimental values, which confirm the overall reliability of the calculated phonon spectra.

The energy dependence of Grüneisen parameter, $\Gamma$, is calculated from the pressure dependence of phonon spectra in the entire Brillouin zone. The calculated dispersion relation at ambient and 0.5 GPa pressure along various high symmetry directions in the Brillouin zone for all three compounds is shown in Fig 7. The computed phonon dispersion relation at 0.5 GPa shows (Fig 7) that transverse acoustic modes show large softening in CuCl and least in CuI. The optic modes in all the compounds are found to harden with pressure.

The maximum negative value (Fig 8) of $\Gamma$ is -3.5, -1.7 and -0.90 for CuCl, CuBr and CuI respectively. It can be seen that low energy modes of energy 4-5 meV have maximum negative value of $\Gamma$. The calculated phonon dispersion relation shows that these modes are mainly transverse acoustic (TA) modes. We find that phonons of energy up to about 6-7 meV show negative $\Gamma$. The volume thermal expansion coefficient, $\alpha_v$, is calculated using the relation (1) for all three compounds and shown in Fig 9. The maximum negative value of $\alpha_v$ in CuCl is $-12.5 \times 10^{-6}$ K$^{-1}$ at 34 K, while in CuBr the value of maximum $\alpha_v$ is $-4.5 \times 10^{-6}$ K$^{-1}$ at 24 K. CuI showed lowest negative values of $-1.0 \times 10^{-6}$ K$^{-1}$ at 10 K.

The low energy modes in all the three halides have nearly same energy. The shorter Cu-Cl bond in CuCl results in higher energy of the optic modes in CuCl in comparison to other two halides, as seen (Figs. 4, 6) from the phonon dispersion relation and density of states plots. The gap between low energy and high energy modes is largest in CuCl and least in CuI. The low energy modes in all the three halides would start contributing at low temperatures, whereas the high energy modes would contribute, as they are populated, with increase in temperature.

The calculated mode $\Gamma$ in all three compounds shows that the transverse acoustic modes have negative $\Gamma$. However for the optic modes $\Gamma$ is positive. The magnitude of negative $\Gamma$ is largest in CuCl and least in CuI. The competition between the contribution from the low energy and high energy modes and magnitude of their $\Gamma$ values would result in net expansion behaviour. In case of CuCl the $\Gamma$ has large negative values and the contribution from the high energy optic modes with positive $\Gamma$ (~ 4 to 5) would be at high temperature. This leads to negative thermal expansion behaviour below 100 K. In other two halides the low energy modes have less negative $\Gamma$ in comparison to CuCl, whereas the high energy optic modes also start contributing at lower temperatures. The overall NTE coefficient in CuBr and CuI has small negative values below 20 K and 10 K respectively.

The calculated fractional change in volume of the CuX is compared with the available experimental data[61] in Fig 10. The agreement between the calculations and experiments is found to be excellent at low temperature, however at higher temperature there is small deviation between the computed and experimental data especially in case of CuCl. This might be due to higher order anharmonic contributions, which are not included in our calculations. The calculated contribution of phonons of energy E to the volume thermal expansion coefficient at 300 K is shown in Fig 11. We find that the maximum contribution to NTE behavior is from phonons of energy 4-5 meV. The other high energy phonons contribute to positive thermal expansion behavior of the compound.

The mean-squared amplitude ($u^2$) of various atoms in CuX have been computed (Fig 12) using the calculated partial density of states. We found that atoms have large mean square amplitude in CuCl in comparison to other halide (X=Cl, Br I) atoms. Further, to understand the role of atomic vibrations responsible for NTE behavior, we have computed the partial contribution of phonons of energy E to the $u^2$ of various atoms (Fig 13(a)) at T=300 K. We observe that phonons of about 4-5 meV contribute maximum to $u^2$. Further in all the three compounds, Cu atoms have larger $u^2$ values in comparison to X atoms. The ratio of the amplitude of Cu to X is maximum in CuCl and least in CuI. The energy range of the peak in the $u^2$ plot matches (Fig. 13 (a)) very well with the energy of zone boundary TA modes at X and L-points in CuX compounds. The eigenvector of the TA phonons at X-point and L-point are plotted in Fig. 14. These mode involve transverse vibrations of the two sub-lattices of Cu and X-atoms.

The nature of bonds is an important factor to understand the thermal expansion behaviour. Our previous work on $M_2O$[62] shows that open structure and ionic nature of bonding results in large variation in thermal expansion behavior of the compounds. The Born effective charges have been calculated (Table III) to understand the nature of bonding. CuBr has somewhat larger magnitude of charge (1.22) that CuCl (1.12) and CuI (1.12). Overall, the Born effective charges seems to be nearly same in all the three compounds, which indicates similar nature of bonding. Further the dielectric constant values (Table III) in all the three compounds are also found to be nearly same.

**D. Negative thermal expansion in CuBr and CuI at high pressures**

The calculated cubic cell length in CuCl is 5.414 Å while that in CuBr and CuI is 5.685 Å and 6.044 Å respectively. The Cu-X bond length in these compounds follow Cu-Cl (2.34 Å) < Cu-Br (2.46 Å) < Cu-I (2.62 Å). The difference in bond lengths at ambient pressure may be responsible for the difference in thermal expansion behavior (Fig. 9) among CuX compounds. In order to investigate the effect of bond length on negative thermal expansion behavior we have calculated the thermal expansion behavior of CuBr and CuI at high pressure *i.e.* lower volume. We find that transverse acoustic branch soften with pressure. While at ambient condition CuBr and CuI show positive thermal expansion behavior, at lower volume (CuBr, a=5.58 Å and P=4.2 GPa; CuI, a=5.76 Å and P=9.8 GPa) the compounds exhibit negative thermal expansion behavior. The calculated dispersion relation at lower volume in CuBr and CuI (Fig 7) shows that the transverse acoustic branch softens significantly, while other branches shift towards high energy. Mainly the transverse acoustic modes in CuBr and CuI shift to lower energies (Fig. 13(b)) in comparison to that at ambient pressure. The large softening of transverse phonon modes in turn may lead to negative thermal expansion in CuBr and CuI at lower volume. The calculations (Fig 8) show that at lower volume the low energy phonons below 8 meV have large negative $\Gamma$ values comparable to CuCl and are responsible for NTE behavior (Fig. 9). These analyses suggest that the magnitude of negative thermal expansion in these compounds behavior is largely dependent on volume.

**V. Conclusions**

We have reported the measurement of temperature dependence of phonon spectra and ab-initio studies on the cuprous halides. The calculated phonon spectra and thermal expansion behavior are in agreement with the available experimental data. The difference in the thermal expansion among these compounds mainly arises from the difference in unit cell volume of these compounds. We show that on reduction of volume, the transverse-acoustic modes of Cu$_4$X shift to lower energies. We predict larger negative Grüneisen parameters and consequently negative thermal expansion in all the three compounds at high pressures.

**Acknowledgements**

S. L. Chaplot would like to thank the Department of Atomic Energy, India for the award of Raja Ramanna Fellowship.

TABLE I. The calculated and experimental[59] lattice parameter of CuX (X=Cl, Br, I).

| Compound | Expt. (Å) | GGA (Å) |
|---|---|---|
| CuCl | 5.414 | 5.432 |
| CuBr | 5.685 | 5.726 |
| CuI | 6.044 | 6.083 |

TABLE II. The computed and experimental[60] elastic constants and Bulk modulus in GPa units for CuX (X=Cl, Br, I).

| Compound | $C_{11}$ | $C_{44}$ | $C_{12}$ | $B$ | $B$ (Expt.) |
|---|---|---|---|---|---|
| CuCl | 57.6 | 16.6 | 45.5 | 49.5 | 38.1 |
| CuBr | 53.0 | 18.3 | 39.0 | 43.7 | 36.6 |
| CuI | 50.8 | 21.4 | 33.2 | 39.1 | 36.6 |

TABLE III  The computed dielectric constants and Born effective charge of various atom in CuX (X=Cl, Br, I).

| Compound | Dielectric constant ($\varepsilon$) | Born effective charges | |
|---|---|---|---|
| | | Cu | X |
| CuCl | 7.3 | 1.12 | -1.12 |
| CuBr | 7.4 | 1.22 | -1.22 |
| CuI | 6.9 | 1.12 | -1.12 |

FIG. 1 (Color online) The crystal structure of CuX (X=Cl, Br and I). Key: Cu, blue sphere; X, green sphere

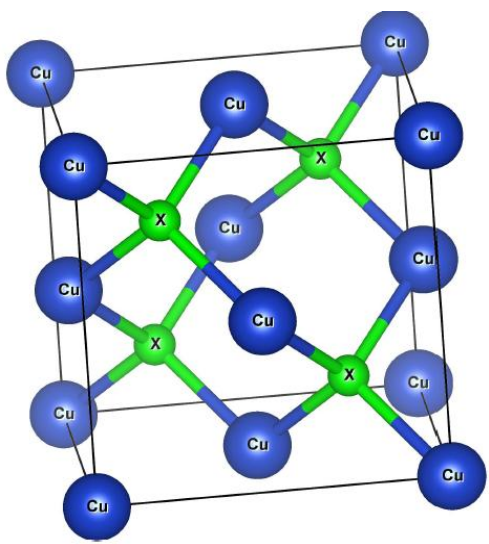

FIG. 2 (Color online) The measured neutron inelastic spectra for CuX (Cl, Br and I) at 373 K, 473 K and 573 K.

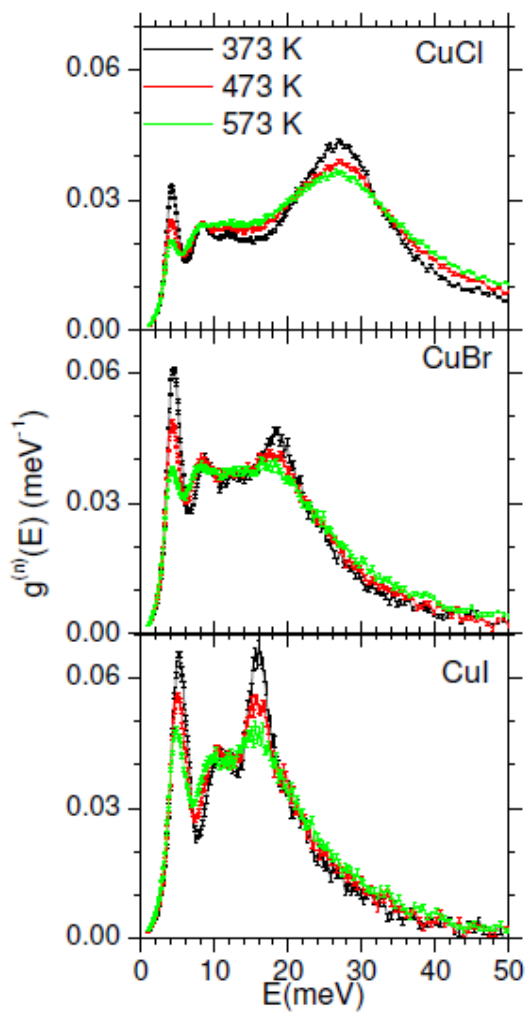

FIG. 3 (Color online) The calculated partial density of states of various atom in CuX (Cl, Br and I).

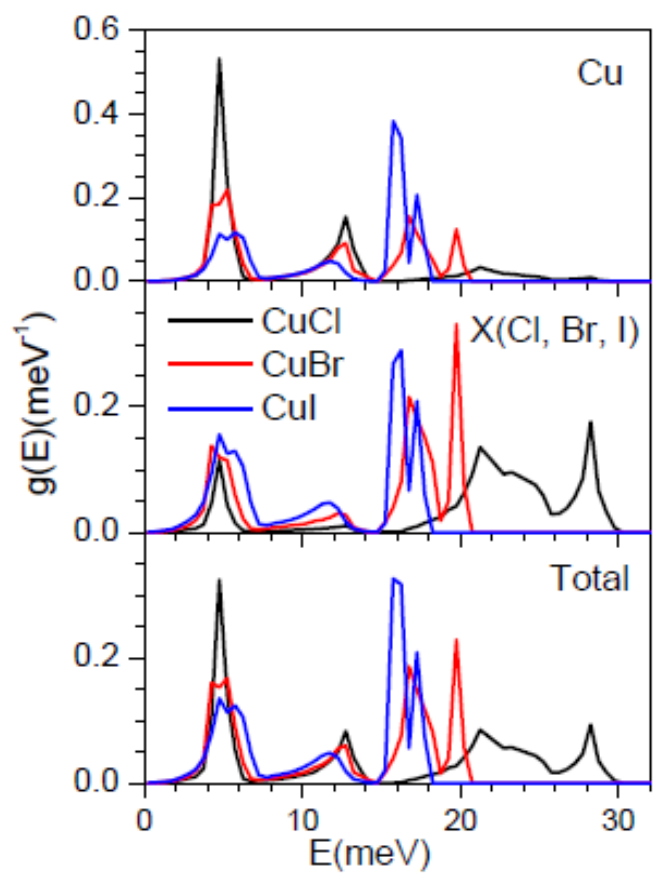

FIG. 4 (Color online) The comparison between the experimental (373 K) and calculated phonon spectra of CuX (Cl, Br and I).

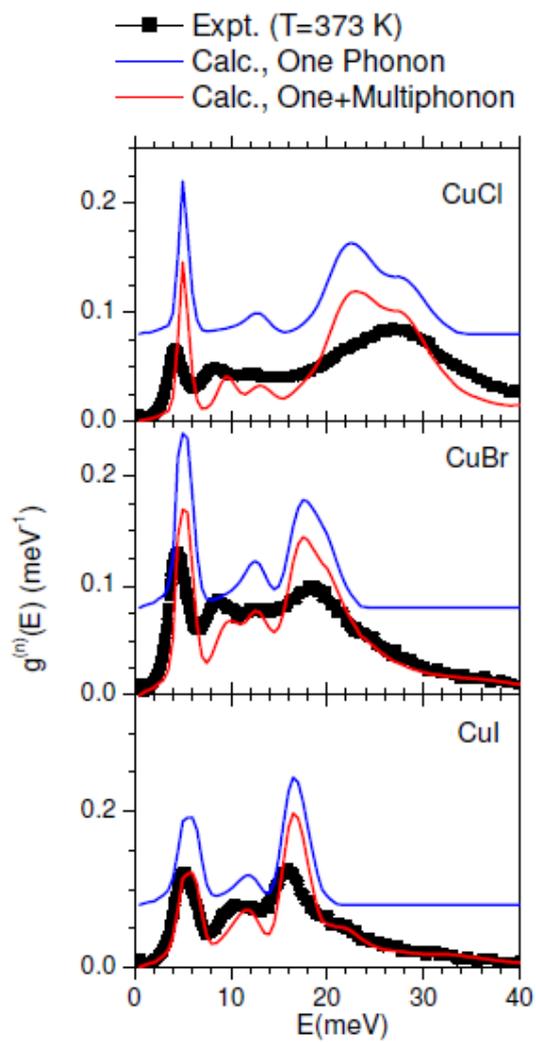

FIG. 5 (Color online) The calculated and experimental[57, 58] specific heat in CuX (Cl, Br and I).

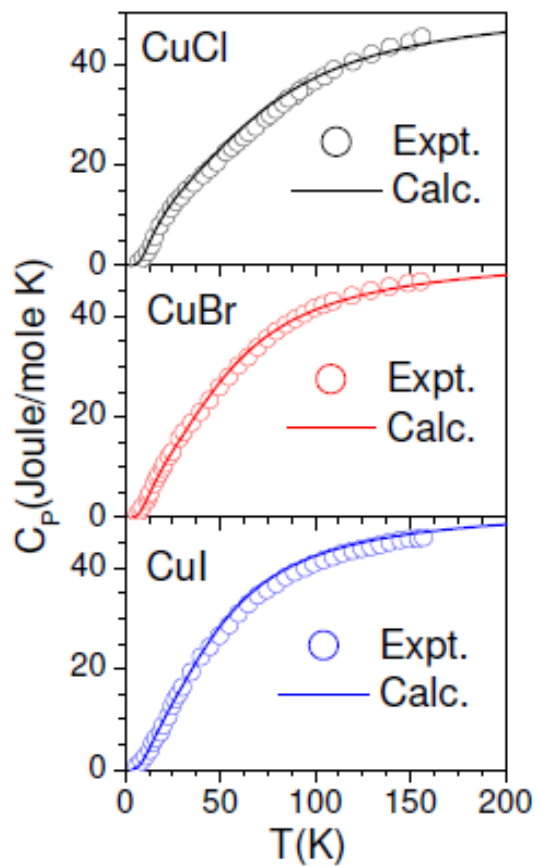

FIG. 6 (Color online) The comparison between calculated phonon dispersion and measured phonon dispersion relation[57, 58] of CuX (X=Cl, Br and I).

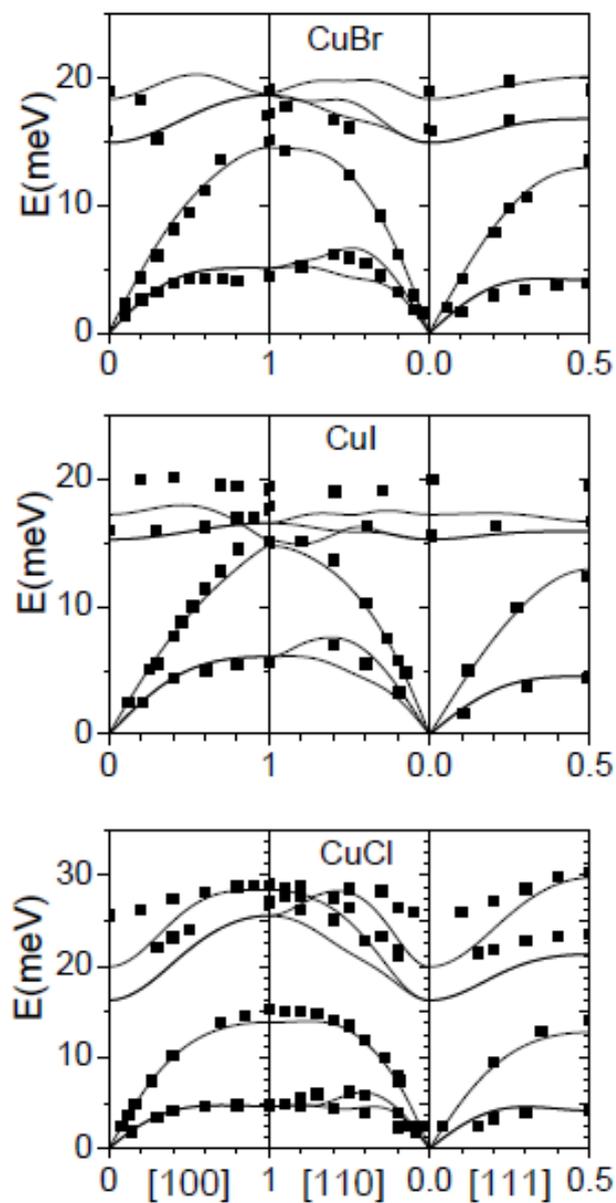

FIG. 7 (Color online) The calculated dispersion relation along various high symmetry directions of CuX (Cl, Br and I) at ambient (black) and high pressures (red). The Bradley-Cracknell notation is used for the high-symmetry points. Γ(0 0 0), X (1 0 0), L( 0.5 0.5 0.5) and W(0.5 1 0) in the notation of cubic unit cell.

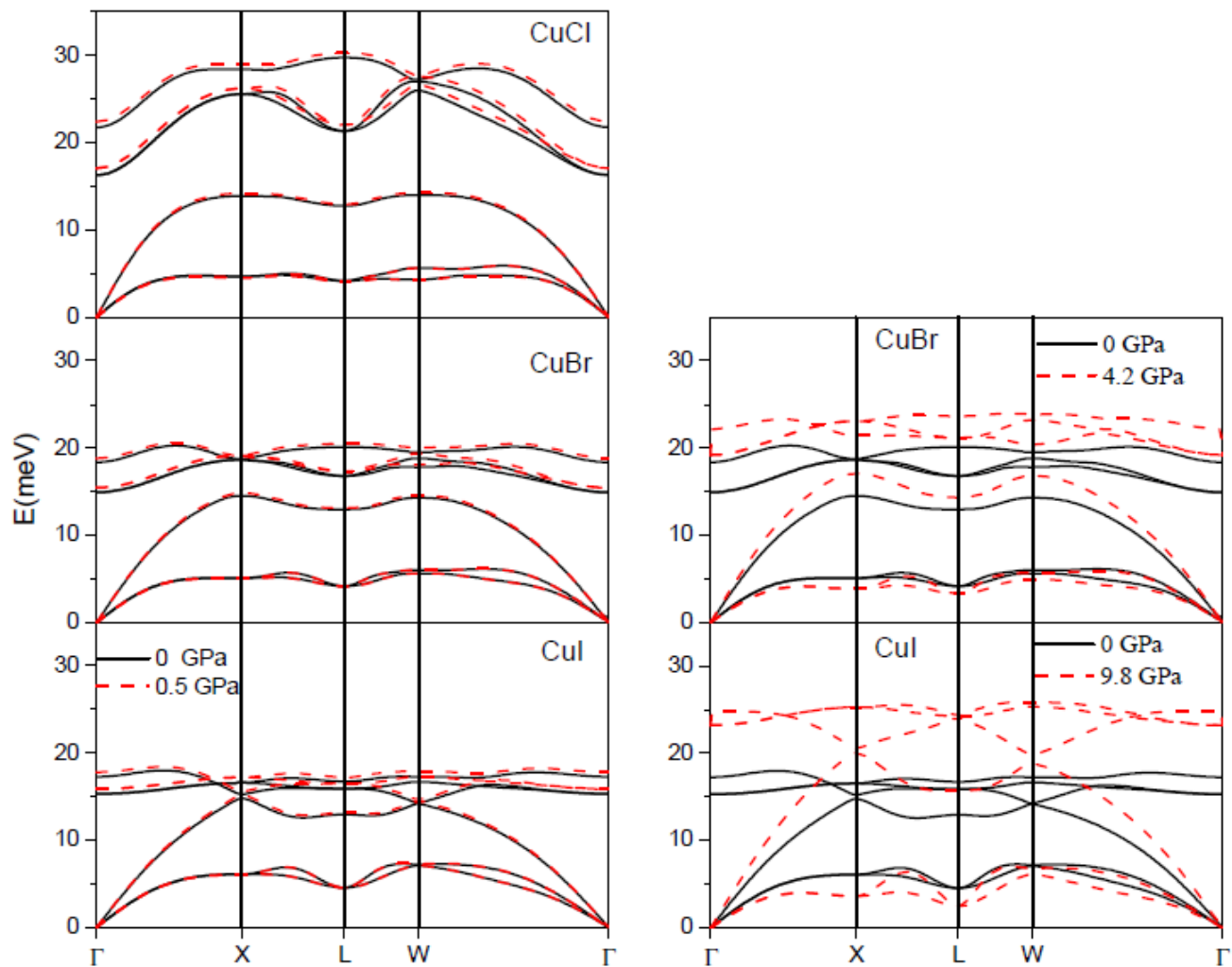

FIG. 8 (Color online) The calculated Grüneisen parameters of CuX (Cl, Br and I) as obtained from pressure dependence of phonon frequency.

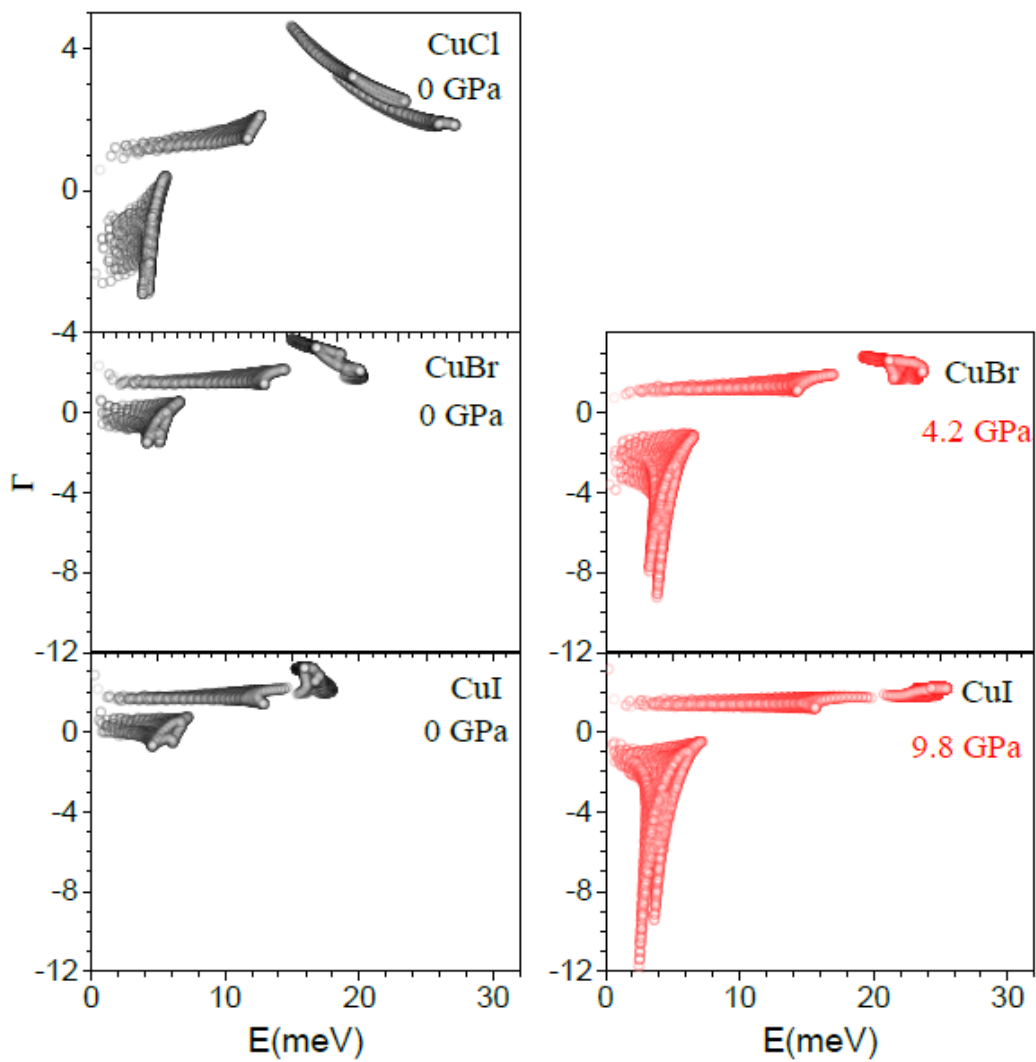

FIG. 9 (Color online) The calculated volume thermal expansion coefficient as a function of temperature of CuX (Cl, Br, I) at (a) ambient pressure and (b) high pressures.

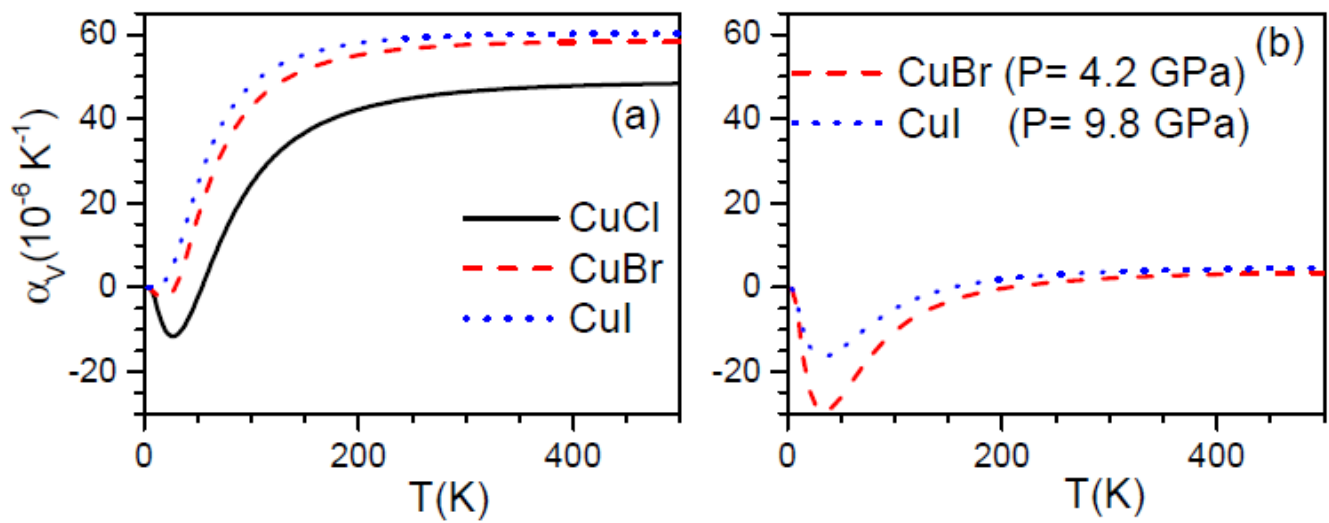

FIG. 10 (Color online) The calculated and experimental[32-34] thermal expansion behavior of CuX (Cl, Br, I).

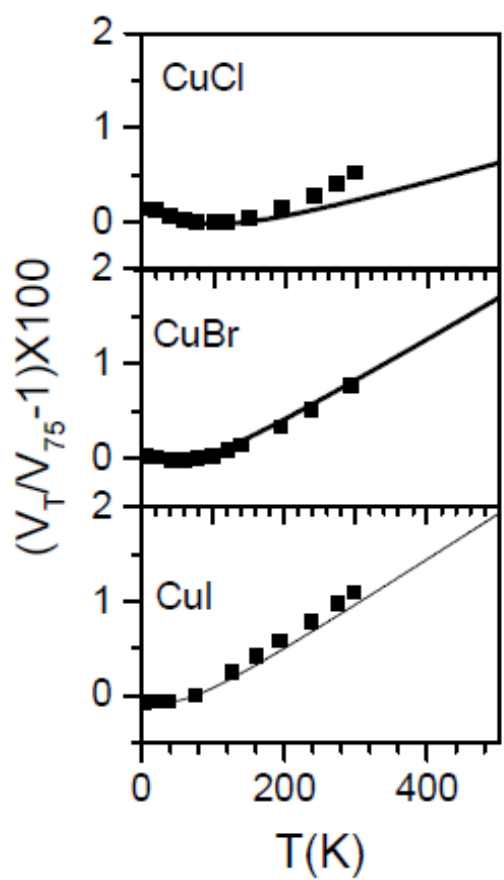

FIG. 11 (Color online) The contribution of phonon of energy E to the volume thermal expansion at T=300 K in CuX (X= Cl, Br, I).

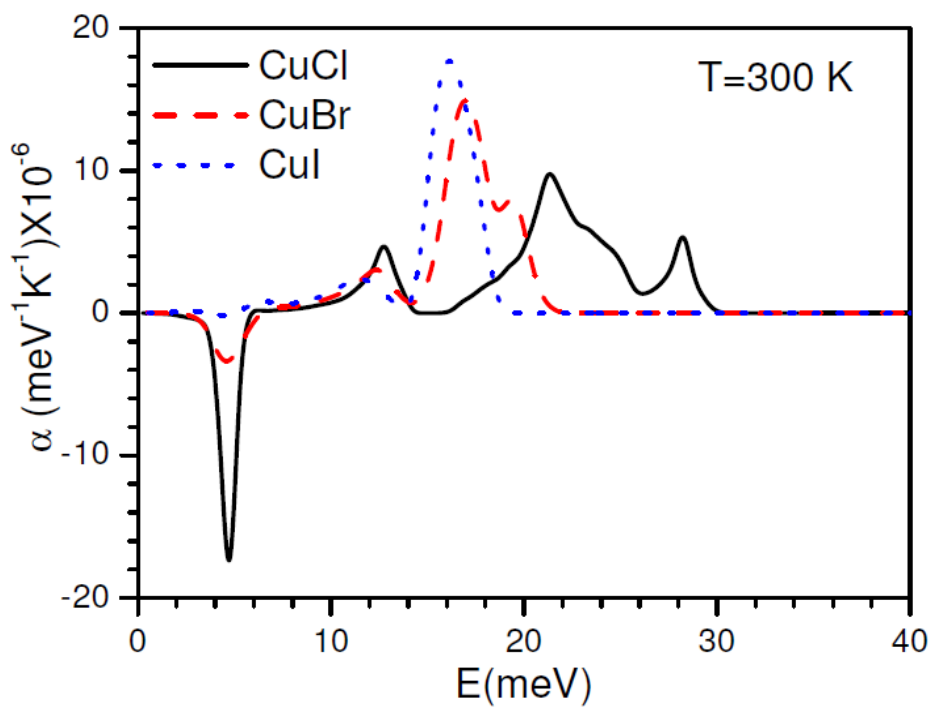

FIG. 12 (Color online) The calculated mean square amplitude of various atom in CuX (Cl, Br and I) as a function of temperature.

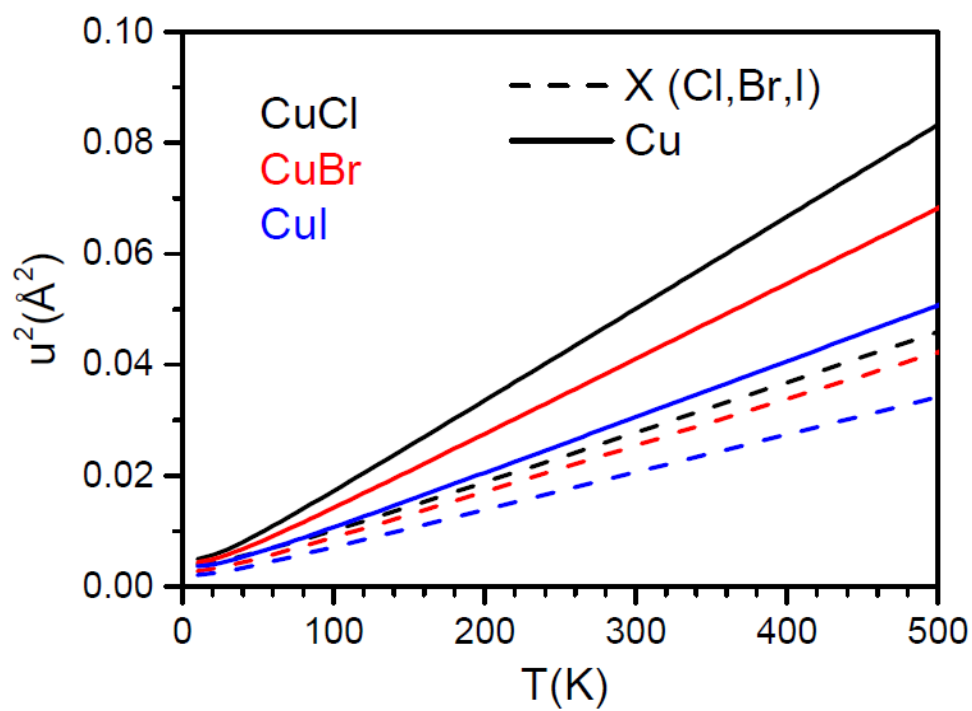

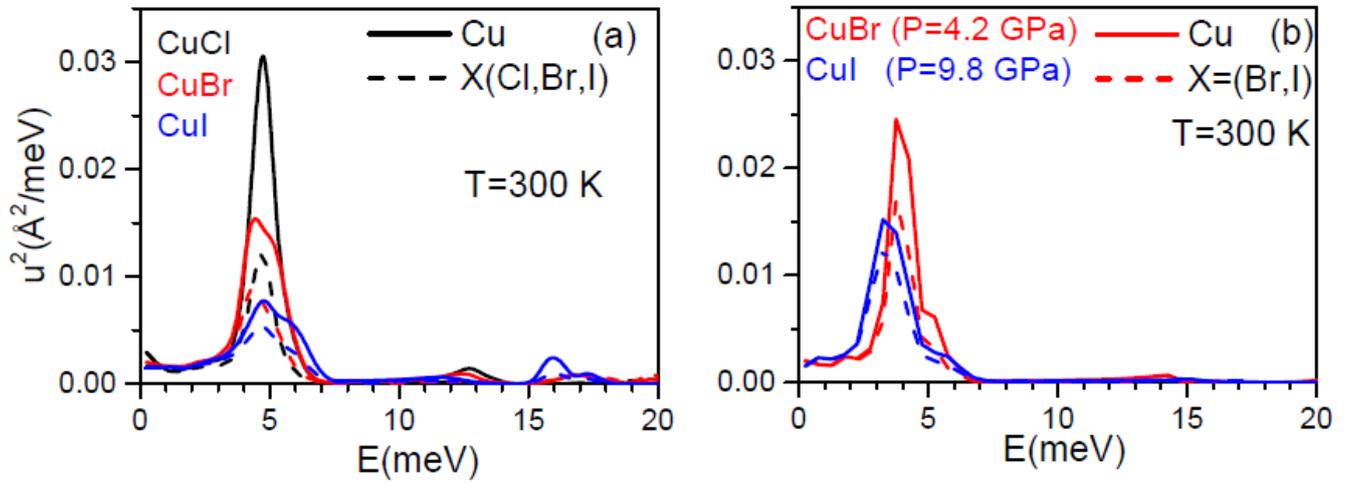

FIG. 13 (Color online) The contribution of phonon mode of energy E in mean square displacement at T=300 K at (a) ambient pressure and (b) high pressures. The full and dashed lines correspond to Cu and X (=Cl, Br, I) atoms.

FIG. 14 (Color online) The eigen vector of various phonon modes in CuX (X=Cl, Br and I). The relative displacements of various atoms are as follows:
X-point mode: (Cu:0.117, Cl:0.060 in CuCl), (Cu:0.106, Br:0.061 in CuBr) and (Cu:0.092, I:0.061 in CuI)
L-point mode: (Cu:0.118, Cl:0.058 in CuCl), (Cu:0.107  Br:0.059 in CuBr) and (Cu:0.093, I:0.060 in CuI)
Key: Cu, blue sphere; X, green sphere

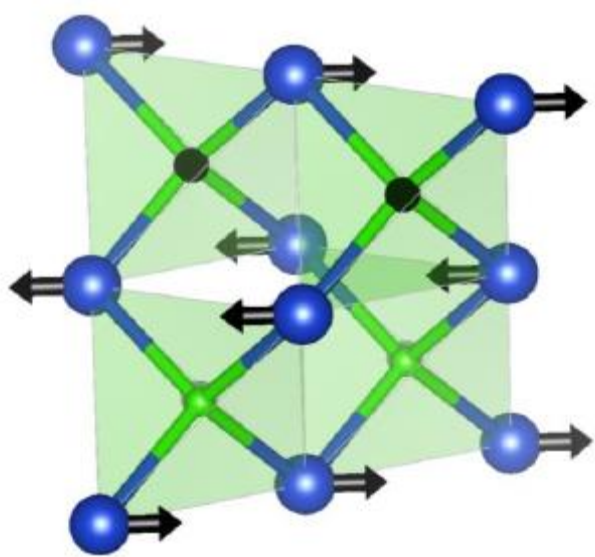
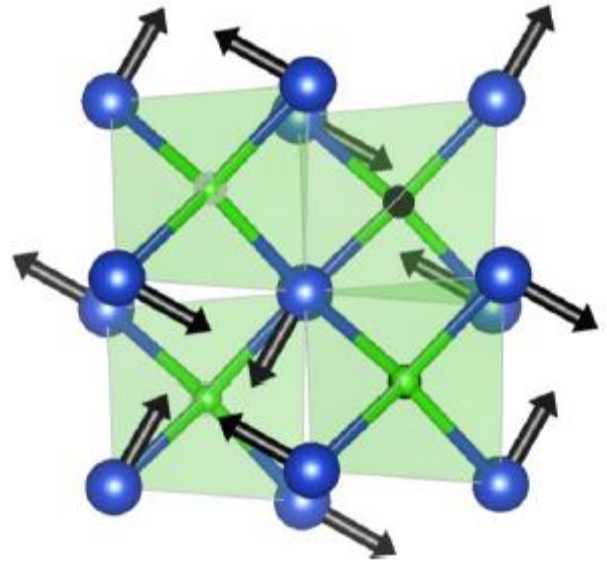

TA Phonon mode at X (1 0 0)         TA Phonon mode at L (1/2 1/2 1/2)